\documentclass[a4paper,fleqn,usenatbib]{mnras}

\usepackage{newtxtext,newtxmath}

\usepackage[T1]{fontenc}
\usepackage{ae,aecompl}
\usepackage{color}

\usepackage{graphicx}	
\usepackage{amsmath}	
\usepackage{amssymb}	

\title[ArH$^+$ dissociative recombination]{Theoretical study of ArH$^+$ dissociative recombination and electron-impact vibrational excitation}

\author[A.~Abdoulanziz et al.]{
A.~Abdoulanziz,$^{1}$
F.~Colboc,$^{1}$
D.~A.~Little,$^{2}$
Y.~Moulane,$^{3,4}$
J.~Zs.~Mezei,$^{1,5,6}$
E.~Roueff,$^7$
\newauthor
J.~Tennyson,$^2$
I.~F.~Schneider$^{1,8}$ and
V.~Laporta$^{1,2}$\thanks{E-mail: vincenzo.laporta@univ-lehavre.fr (VL)}
\\
$^1$Laboratoire Ondes et Milieux Complexes, CNRS--Universit{\'{e}} du Havre--Normandie Universit{\'{e}}, 76058 Le Havre, France\\
$^2$Department of Physics and Astronomy, University College London, London WC1E 6BT, UK\\
$^3$Oukaimden Observatory, High Energy Physics and Astrophysics Laboratory, Cadi Ayyad University, Marrakech, Morocco\\
$^4$Space sciences, Technologies \& Astrophysics Research Institute, University of Li\`{e}ge, Li\`{e}ge, Belgium\\
$^5$Laboratoire des Sciences des Proc\'ed\'es et des Mat\'eriaux, CNRS$-$Universit\'e Paris 13$-$USPC, 93430 Villetaneuse, France\\
$^6$Institute of Nuclear Research, Hungarian Academy of Sciences, Debrecen, Hungary\\
$^7$Sorbonne Universit{\'{e}}, Observatoire de Paris, Universit{\'{e}} PSL, CNRS, LERMA, F-92190,  Meudon, France \\
$^8$Laboratoire Aim\'{e}-Cotton, CNRS$-$Universit\'e Paris-Sud$-$ENS Cachan$-$Universit\'e Paris-Saclay, 91405 Orsay, France
}

\date{Accepted XXX. Received YYY; in original form ZZZ}

\pubyear{2018}

\begin{document}
\label{firstpage}
\pagerange{\pageref{firstpage}--\pageref{lastpage}}
\maketitle

\begin{abstract}
Cross sections are presented for dissociative recombination and
electron-impact vibrational excitation of the ArH$^+$ molecular ion
  at electron energies appropriate for the interstellar environment.
  The R-matrix method is employed to determine the molecular structure
  data, \textit{i.e.} the position and width of the resonance states.
  The cross sections and the corresponding Maxwellian rate
  coefficients are computed using a method based on the Multichannel
  Quantum Defect Theory. The main result of the paper is the very low dissociative recombination rate found at temperatures below 1000K. This is in agreement with the previous upper limit measurement in merged beams and offers a realistic explanation to the presence of ArH$^+$ in exotic interstellar conditions.
\end{abstract}




\section{Introduction}

The presence of the ArH$^+$ molecular cation, argonium, in interstellar medium (ISM) was
reported for the first time by \citet{Barlow2013}, who detected $^{36}$ArH$^+$
617.525 GHz ($J= 1 -0$) and 1234.603 GHz ($J= 2-1$) emission lines in spectra from the Crab Nebula using
the data from Herschel mission. 
That supernova remnant is known to contain both molecular
hydrogen and regions of enhanced ionized argon emission.  After this first noble
gas molecular ion detection, \cite{Schilke2014} realized that the still unidentified
absorption transition at 617.5 GHz observed in diffuse gas toward several
sources such as Sg B2, and various PRISMA sources (W31C, W49N, W51e, \ldots), was
in fact due to argonium with $^{36}$Ar.  Moreover, features of $^{38}$ArH$^+$ were
subsequently found in Sg B2(M) as well and, consequently,  Schilke {\it et al.} suggested that argonium is 
ubiquitous in the ISM. More recently, \cite{Muller2015} made extragalactic
detections of the $^{36}$Ar and $^{38}$Ar isotopologues  of argonium
through absorption studies of a foreground
galaxy at $z= 0.89$ along two different lines of sight toward PKS 1830-211
within the band 7 of the ALMA interferometer, including the corresponding
redshifted transitions.

The possible formation/destruction processes linked to ArH$^+$ are discussed by
\citet{Neufeld2016} who emphasized that  ArH$^+$  is a good tracer of the almost
purely atomic diffuse ISM in the Milky Way.  
However, an important missing piece of information remains the unknown value 
of the dissociative recombination rate coefficient of that molecular ion.
An upper limit of 10$^{-9}$ cm$^3$ s$^{-1}$  for electron
collision energies below about 2 eV was reported by \citet{Mitchell2005}
who performed a storage ring measurement. Mitchell {\it et al.}  
also gave the corresponding theoretical potential curves. That upper  limit
value is adopted in the presently available astrochemical 
models for galactic diffuse clouds \citep{Neufeld2016} whereas \citet{Priestley2017} introduce a lower value
(10$^{-11}$ cm$^3$ s$^{-1}$) to interpret the Crab nebula observations.
 Photodissociation of ArH$^+$, another potential
destruction mechanism, was studied theoretically by \citet{Alekseyev2007}
and the corresponding photodissociation rate was shown to be moderate,
\textit{i.e.} 9.9 10$^{-12}$  s$^{-1}$ in the unshielded mean ultraviolet
interstellar radiation field \citep{Roueff2014, Schilke2014}. In addition to these, the rotational excitation due to
electron impact has been studied by \citet{HamiltonFaureTennyson2016}.

In this paper, we investigate theoretically the dissociative recombination (DR) 
process of ArH$^+$  through  \emph{ab initio} methods, including the
dependence on the  vibrational excitation of the target molecular ion and, in the same energy range, the
competitive  process of vibrational excitation  (VE) by electron impact, \textit{i.e.}: 
\begin{align}
e(\varepsilon)+\textrm{ArH}^+(\textrm{X}\,^1\Sigma^+,v^+) \to & \textrm{Ar} +
\textrm{H}\,, & (\textrm{DR}) \label{eq:DRprocess}
\\
e(\varepsilon) + \textrm{ArH}^+(\textrm{X}\,^1\Sigma^+,v^+) \to &
\textrm{ArH}^+(\textrm{X}\,^1\Sigma^+,w^+) + e\,, & (\textrm{VE})
\label{eq:VEprocess}
\end{align}
where $\varepsilon$ is the incident electron energy, $v^+$ and $w^+$ represent
the initial and final vibrational quantum numbers respectively corresponding to the  ground electronic state $\textrm{X}\,^1\Sigma^+$
of ArH$^+$. 

The manuscript is organized as follows: in Section \ref{sec:PEC} the theoretical
model used to characterize the ArH$^{**}$ resonant states is presented and in
Section \ref{sec:results} the results concerning the cross sections and the
corresponding rate coefficients are discussed. Finally the conclusions, in
Section \ref{sec:concl}, close the paper.

\section{Theoretical model}\label{sec:PEC} 

A theoretical study of the ArH$^+$ electronic excited states was
performed by \citet{B501400J}; \citet{Jungen1481}, and more recently 
\citet{B511864F}, explored  ArH Rydberg states.

In the present work, \textit{ab initio}  ArH$^+$ calculations were performed using
MOLPRO and an aug-cc-pVQZ (AVQZ) Gaussian type orbital (GTO) basis set
at the complete active space (CAS) self-consistent field (SCF) level of theory.
These calculations provided input orbitals for the electron-ion scattering calculations.
All calculations were performed in C$_{2v}$ symmetry, which is the highest
allowed by MOLPRO and the polyatomic R-matrix code for an asymmetric linear molecule.

The potential energy curves and the widths for the ArH$^{**}$ resonant states were calculated using the R-matrix method \citep{Tennyson_PR_2010} as implemented in UKRMol code \citep{Carr2012}. The general approach follows closely the treatment of N$_2^{**}$ by \citet{0953-4075-47-10-105204} which provided the input for N$_2^+$ DR calculations \citep{Little-PRA-2014}. The ArH$^+$ target states were represented using the AVQZ GTO basis set and a CAS in which the Ar 1s$^2$2s$^2$2p$^6$ electrons were frozen and the remaining 8 electrons were  distributed as $(4\sigma,5\sigma,6\sigma,2\pi)^8$.  The $3\pi$ virtual orbital was retained to augment the continuum orbitals in the scattering calculation.

The scattering calculations used an R-matrix sphere of radius 10~a$_0$. Continuum basis functions were represented using GTOs placed at the center of this sphere and contained up to $g$ orbitals ($\ell \leq 4$) \citep{Faure2002224}. Close-coupling calculations built on the target CAS \citep{0953-4075-29-24-024} and an expansion of the 8 lowest states of each (C$_{2v}$) symmetry were retained for the outer region calculations. In this latter region, calculations were repeated for the  internuclear separations $2.2<R<15$~a$_0$ and for symmetries corresponding to  $^2\Sigma^+$, $^2\Pi$ and $^2\Delta$ scattering channels.

The outer region calculations explicitly considered the 20 lowest target
states. R-matrices were propagated to 100.1~a$_0$ and then fitted to an asymptotic form. Resonance positions
and widths were determined by automated fitting of the eigenphase sums to a Breit-Wigner form
using program RESON  \citep{Tennyson1984421}. Couplings were determined from the resonance widths $\Gamma$ using the formula:
\begin{equation}
V(R)=\sqrt{\frac{\Gamma(R)}{2\pi}}\,.
\end{equation}

\begin{figure}
\centering
\includegraphics[scale=.3]{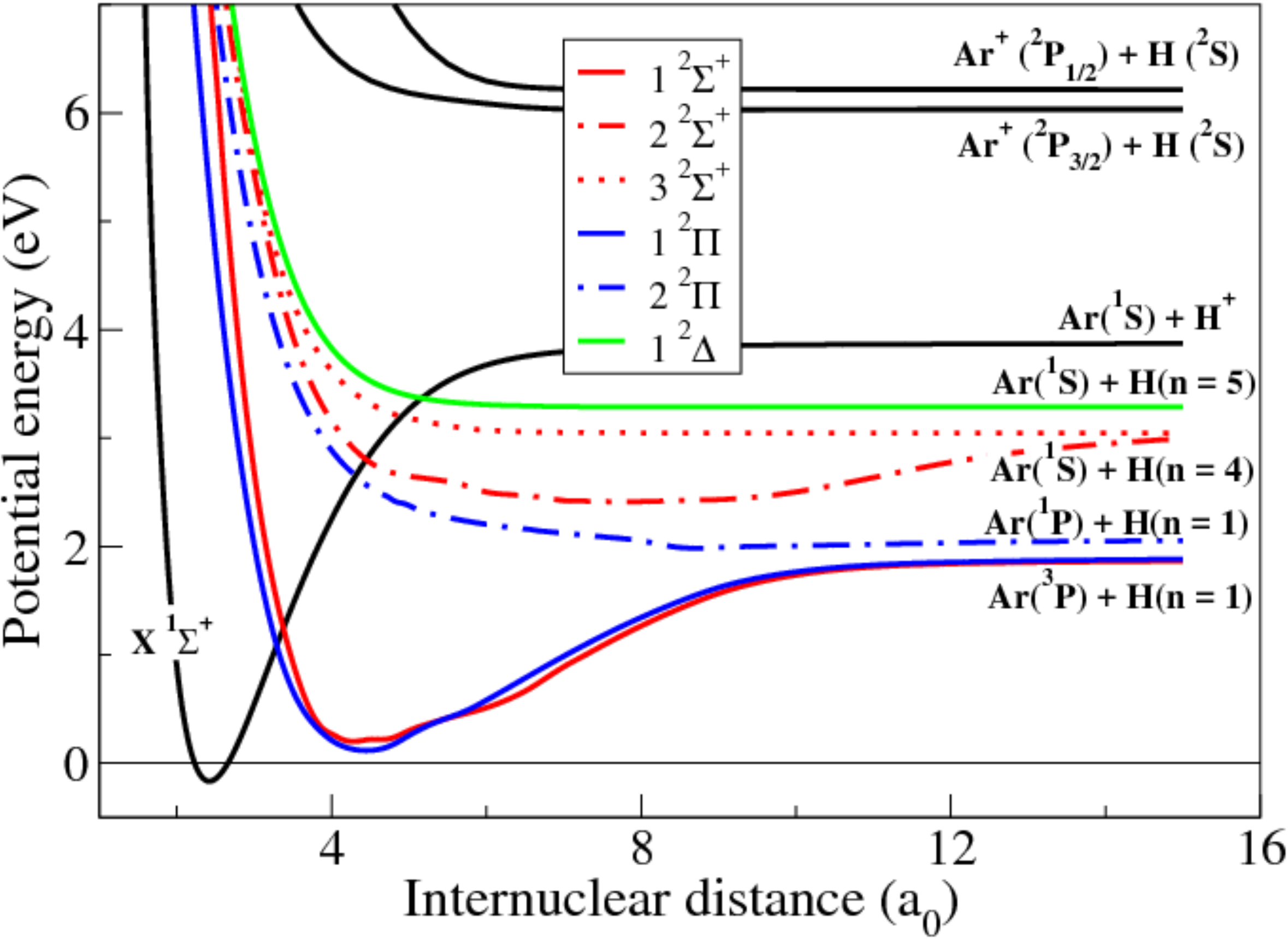}
\includegraphics[scale=.3]{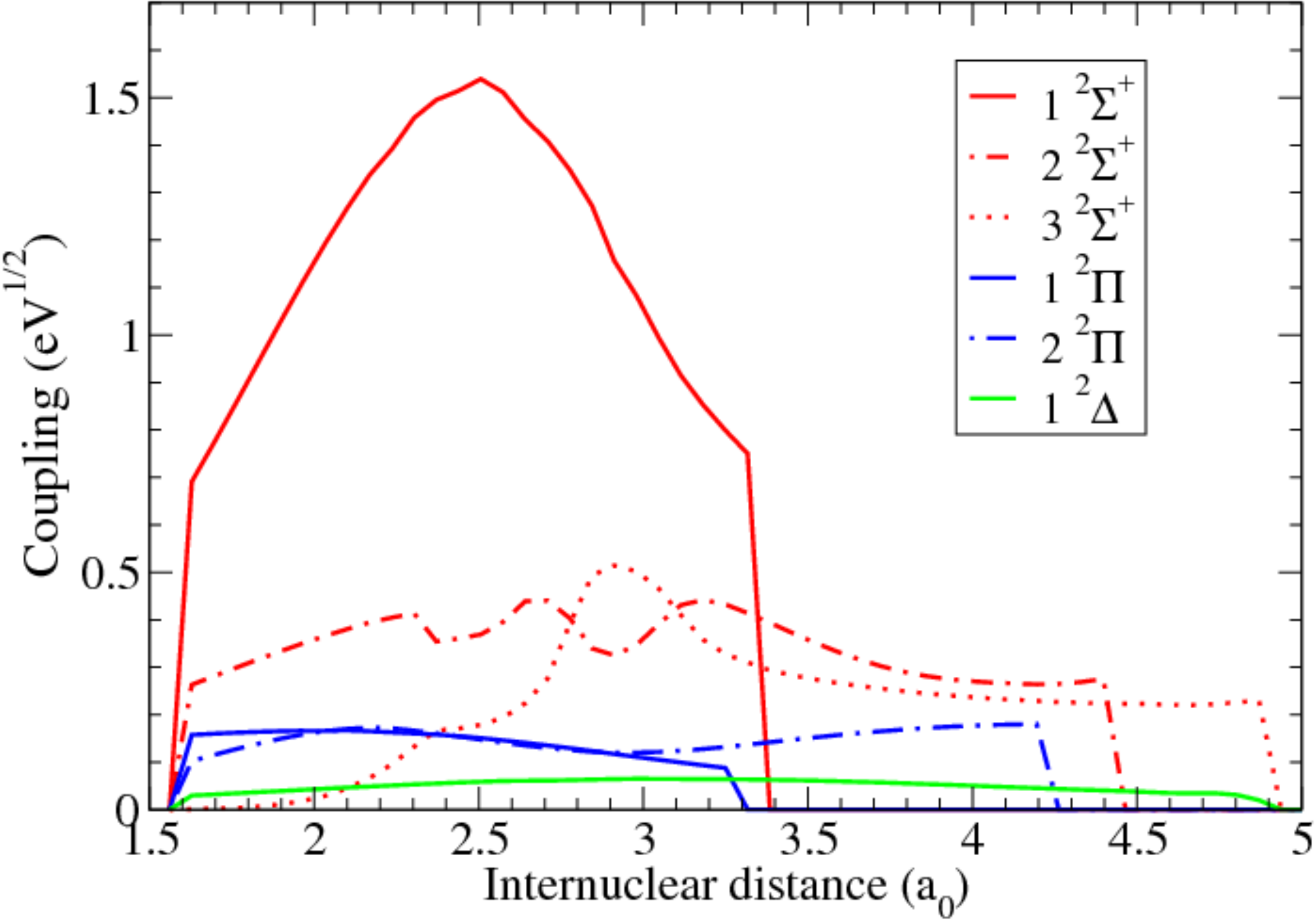}
\includegraphics[scale=.3]{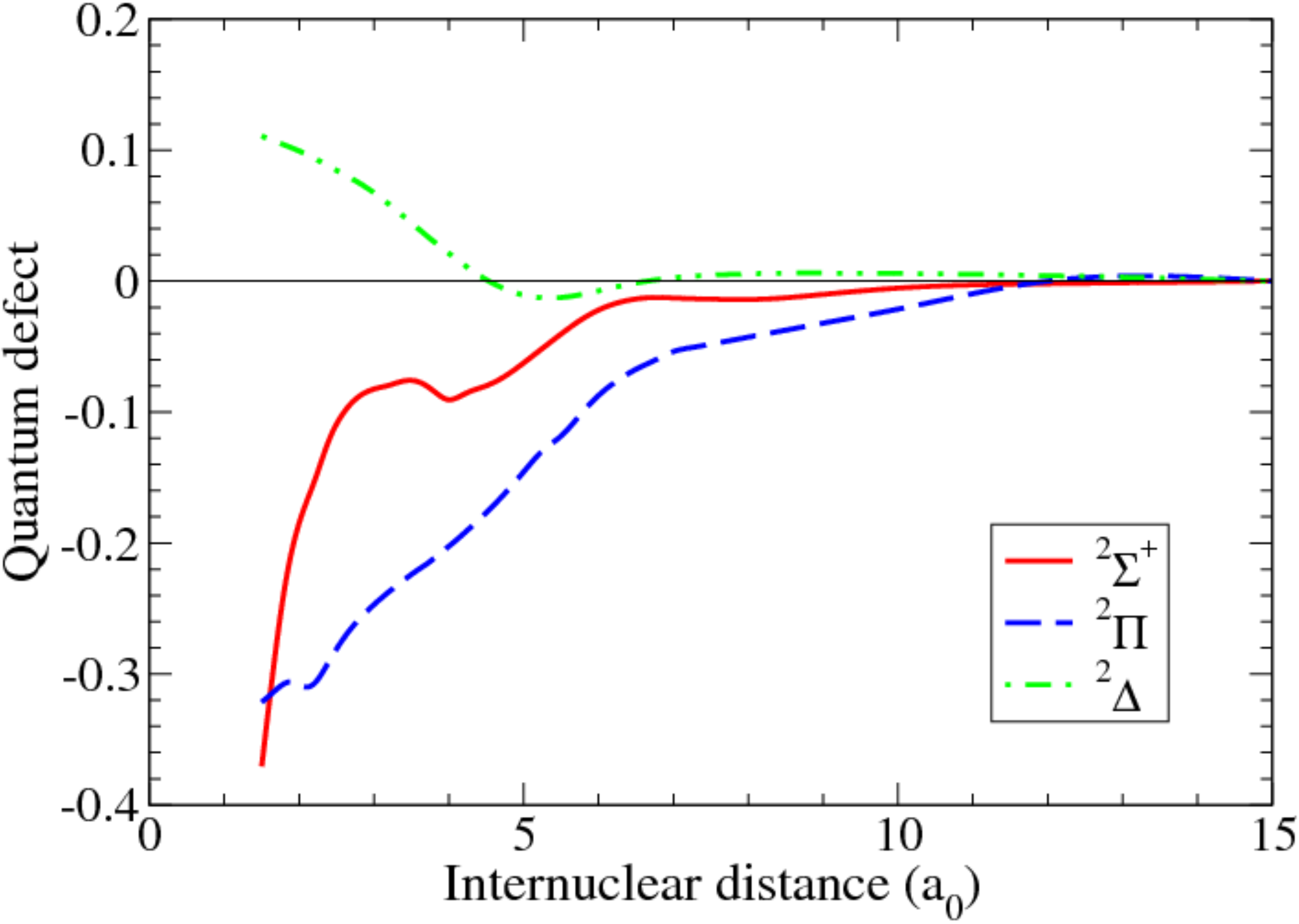}
\caption{Potential energy curves, couplings and quantum defects used in the present calculations. The ArH$^+$ potential curves - ground state, $\textrm{X}\,^1\Sigma^+$, and the lowest excited electronic states - are displayed as black lines. The molecular data sets for the different symmetries of the neutral system are displayed with different colors: $^2\Sigma$  in red, $^2\Pi$  in blue  and $^2\Delta$ in green. \label{fig:pecs}}
\end{figure}

Figure \ref{fig:pecs}  shows the R-matrix results for resonance positions (upper panel), couplings (middle panel) and quantum defect (lower panel). The corresponding molecular data are given in Table \ref{tab:ArH+data}. These data form the input for the  Multichannel Quantum Defect Theory (MQDT) step of the calculations. Linear extrapolation was adopted for the couplings in order to extend the internuclear distances range below 2.2~a$_0$ to 1.6~a$_0$. 

\begin{table}
\centering
\begin{tabular}{cccc}
\hline
$\mu$ (a.u.) & \multicolumn{3}{c}{1791.94} \\
\hline
$R_{eq}$ ($a_0$) & \multicolumn{3}{c}{2.419 (2.419)}  \\
\hline
$D_e$ (eV) & \multicolumn{3}{c}{4.039 (4.025)} \\
\hline
$D_0$ (eV) & \multicolumn{3}{c}{3.8725} \\
\hline\hline
~~~$v^+$~~~ & $~~\epsilon_{v^+}$~(eV)~~ & ~~~$v^+$~~~ & $~~\epsilon_{v^+}$~(eV)~~\\
\hline
0  & 0.000 & 12 & 2.949 \\
1  & 0.321 & 13 & 3.110 \\
2  & 0.627 & 14 & 3.258 \\
3  & 0.919 & 15 & 3.393 \\
4  & 1.197 & 16 & 3.513 \\
5  & 1.461 & 17 & 3.617 \\
6  & 1.712 & 18 & 3.703 \\
7  & 1.949 & 19 & 3.770 \\
8  & 2.174 & 20 & 3.817 \\
9  & 2.387 & 21 & 3.846 \\
10 & 2.587 & 22 & 3.861 \\
11 & 2.774 &    & \\
\hline
\end{tabular}
\caption{Molecular constants (reduced mass, equilibrium distance and dissociating energies) for $^{40}\mathrm{ArH}^+$ in its ground electronic state and the energies of the corresponding vibrational levels. The comparison with the experimental data of \citet{Hotop1998}  given in brakets is reported. \label{tab:ArH+data}}
\end{table}

ArH$^+$ is a closed shell system so no spin-orbit (SO) splitting effects are expected in its ro-vibrational levels. Conversely, SO effects may be important in the non-$\Sigma$ resonances and are well characterized for the Ar asymptotic states. In particular, the Ar($^2\mathrm{P}^0_{3/2}4s$) and Ar($^2\mathrm{P}^0_{1/2}4s$) show SO splittings of 0.075 eV and 0.105 eV, respectively \citep{NIST_ASD}. Our calculations are non-relativistic and therefore neglect SO effects; we assume  the calculated R-matrix resonances converge on the lowest component of the Ar doublets at large internuclear distances. Table \ref{tab:Ar+H_limits} shows the asymptotic limits of the ArH$^{**}$ resonant states considered below.
\begin{table}
\centering
\begin{tabular}{ccl}
\hline
Channel  & Energy (eV) & Symmetries\\
\hline
$\mathrm{Ar}(^3\mathrm{P})+\mathrm{H}(\mathrm{n}=1)$ & -2.00 (-2.05) &  $1\,^2\Sigma^+$, $1\,^2\Pi$\\
$\mathrm{Ar}(^1\mathrm{P})+\mathrm{H}(\mathrm{n}=1)$ & -1.81  (-1.87) &  $2\,^2\Pi$\\
$\mathrm{Ar}(^1\mathrm{S})+\mathrm{H}(\mathrm{n}=4)$ & -0.87 (-0.85) &  $2\,^2\Sigma^+$, $3\,^2\Sigma^+$\\
$\mathrm{Ar}(^1\mathrm{S})+\mathrm{H}(\mathrm{n}=5)$ & -0.58 (-0.54) &  $1\,^2\Delta$\\
\hline
\end{tabular}
\caption{Asymptotic limits of the ArH$^{**}$ resonant  states  relevant for the low-energy impact collisions. The energy is expressed with respect to the asymptotic limit of the ground electronic state of ArH$^+$. The experimental energy values from the NIST database \citep{NIST_ASD} are given in brackets for comparison. \label{tab:Ar+H_limits}}
\end{table}

The MQDT method \citep{0022-3700-13-19-025, doi:10.1063/1.460913, Chakrabarti-PRA-2013, PhysRevA.90.012706, Little-PRA-2014, Epee-MNRAS-2015}  was used to study the processes (\ref{eq:DRprocess}) and (\ref{eq:VEprocess}). Within this approach, the corresponding cross sections are expressed in terms of  S-matrix elements as:
\begin{align}
\sigma_{v^{+}}(\varepsilon) &= \frac{\pi}{4\varepsilon} 
\sum_{sym,\Lambda,l,j}\rho^{sym,\Lambda}\left|  S^{sym,\Lambda}_{d_j,lv^+} \right
|^2\,,\label{eq:DRxsec}
\\
\sigma_{v^{+},w^+}(\varepsilon) &= \frac{\pi}{4\varepsilon} 
\sum_{sym,\Lambda,l,l'}\rho^{sym,\Lambda}\left| S^{sym,\Lambda}_{l'w^+,lv^+} -
\delta_{l,l'}\delta_{v^+,w^+} \right |^2\,,\label{eq:DRxsec}
\end{align}
where the summation is extended over all symmetries ({\it sym}: spin, inversion for homonuclear molecules) of the neutral system, 
projection of the total electronic angular momentum on the internuclear axis $\Lambda$ , and partial waves $l$$/$$l'$ of the incident/scattered electron, and $\rho^{sym,\Lambda}$ is the ratio between  the multiplicities of the neutral system and of the target ion.

The most abundant isotope of argon in the Earth's atmosphere is $^{40}$Ar whereas in the ISM $^{36}$Ar and $^{38}$Ar isotopes are preponderant. In the present work,  we deal with vibrational processes and, due to the small relative variation of the reduced mass from one  isotopologue to an other - as a consequence of the huge atomic mass of the Ar isotopes - we expect these effects to be negligible. In order to verify this, we performed calculations for different isotopologues of ArH$^+$ and the relative difference between the rate coefficients  was found to be below $1\,\%$.

\section{Results and discussion} \label{sec:results}

\begin{figure}
\centering
\includegraphics[scale=.3]{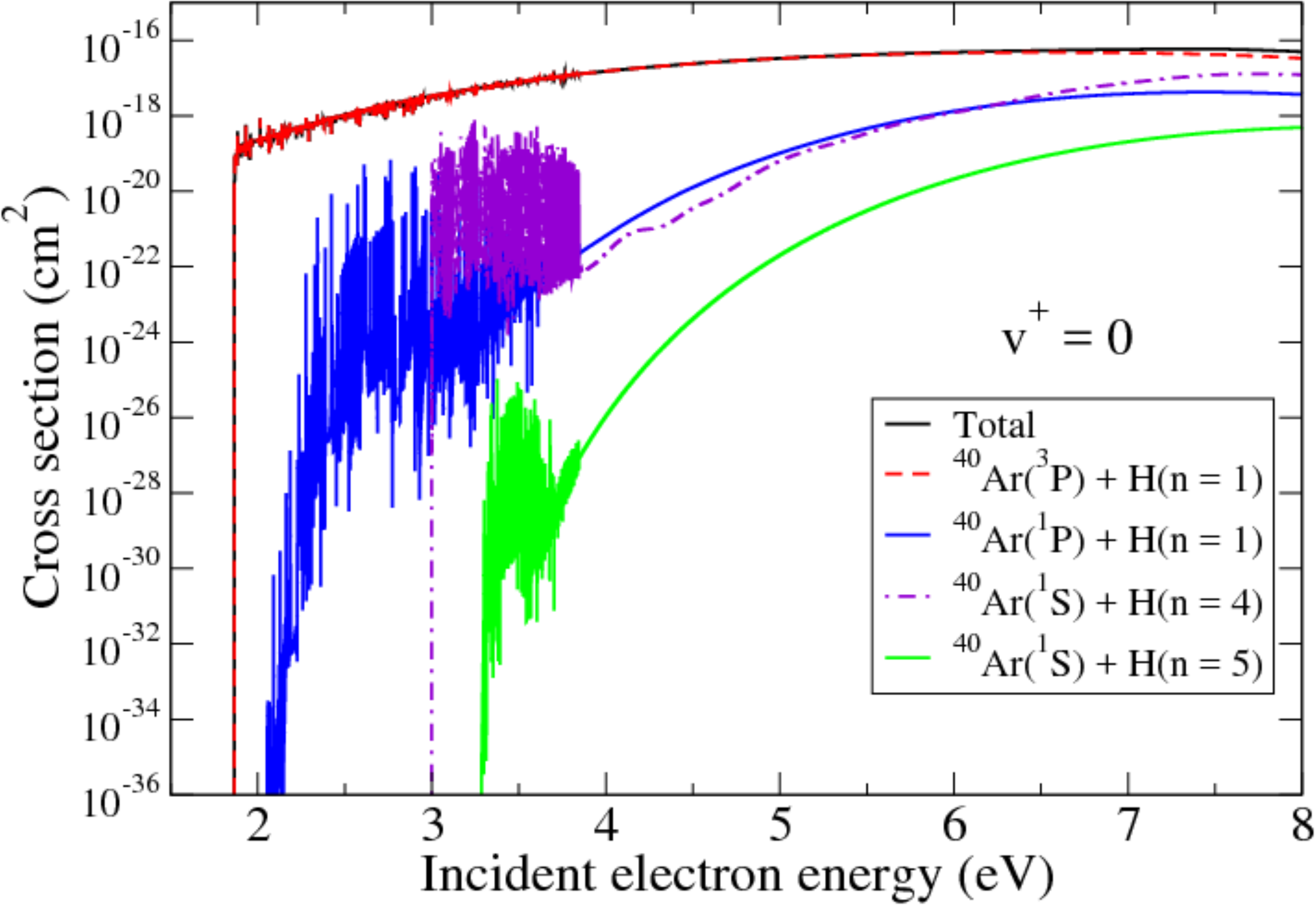}
\caption{Dissociative recombination of vibrationally relaxed  ArH$^+$. Broken colored lines: The  contributions coming from all the dissociative states having the same asymptotic atomic limit. Solid black line (partially hidden by the red curve): Total cross section coming from the sum over all the available dissociative states.
\label{fig:xsecDR_symm}}
\end{figure}

Figure \ref{fig:xsecDR_symm} displays the DR cross sections for ArH$^+$ $v^+=0$, namely the total one
and  the partial contributions corresponding to the asymptotic channels of resonant states.
It can be noted that the main contribution arises from the $\mathrm{Ar}(^3\mathrm{P})+\mathrm{H}(n=1)$ channel. One reason for this is that, as shown in Table \ref{tab:Ar+H_limits}, this exit channel gathers
contributions coming from two states - $1\,^2\Sigma^+$ and $1\,^2\Pi$ - instead of one state, as is the case of the exit channels  
$\mathrm{Ar}(^1\mathrm{P})+\mathrm{H}(n=1)$
and
$\mathrm{Ar}(^1\mathrm{S})+\mathrm{H}(n=5)$
. One can argue that - as shown in Table  \ref{tab:Ar+H_limits} - the channel
$\mathrm{Ar}(^1\mathrm{S})+\mathrm{H}(n=4)$ 
is the asymptotic limit of two states, as the
$\mathrm{Ar}(^3\mathrm{P})+\mathrm{H}(n=1)$ 
one. However, the coupling of the $1\,^2\Sigma^+$ state with the electron/ion continuum (see Fig.~\ref{fig:pecs}) 
is about three times larger than the other ones.

\begin{figure*}
\centering
\includegraphics[scale=.3]{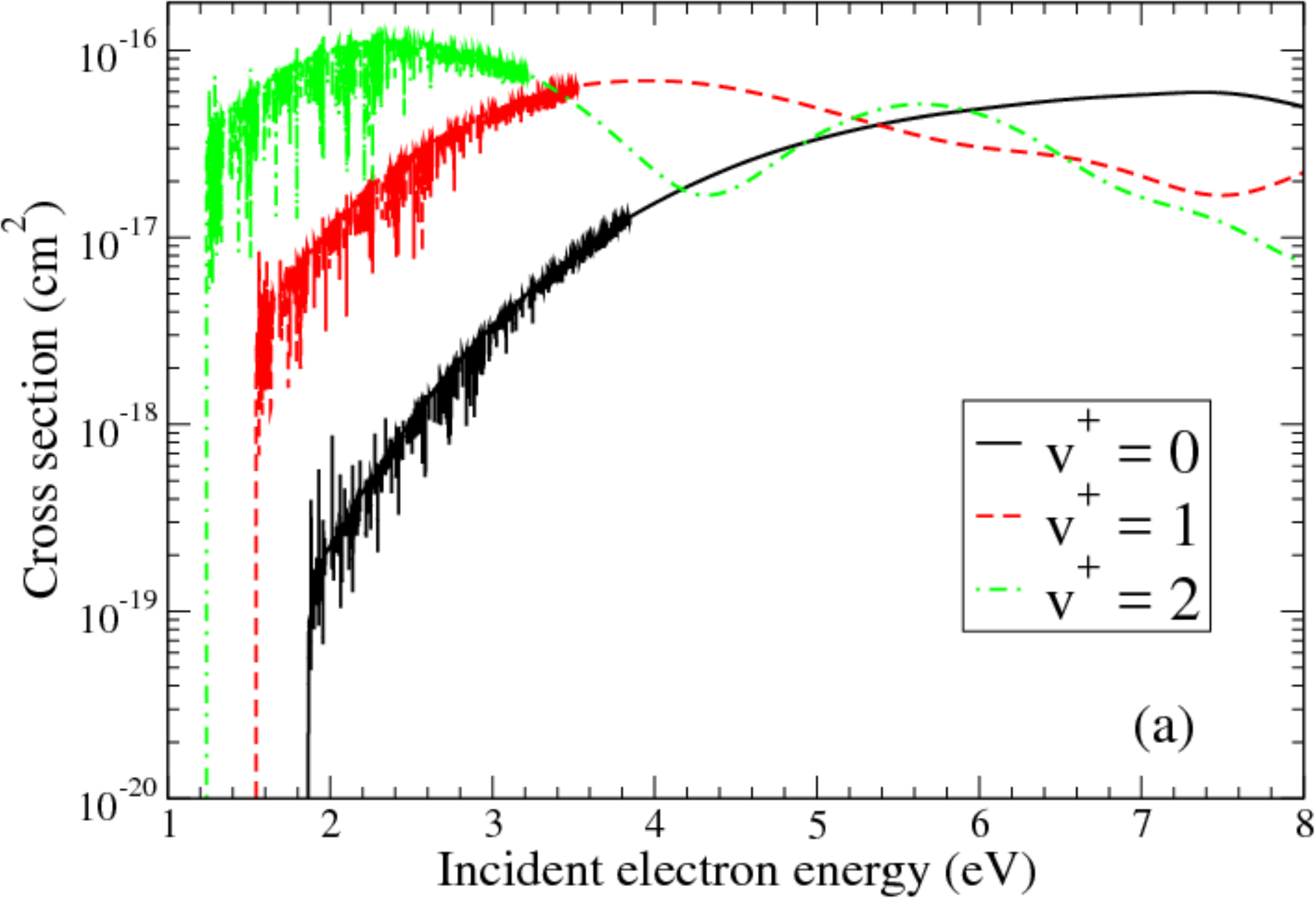}\hspace{1cm} \includegraphics[scale=.3]{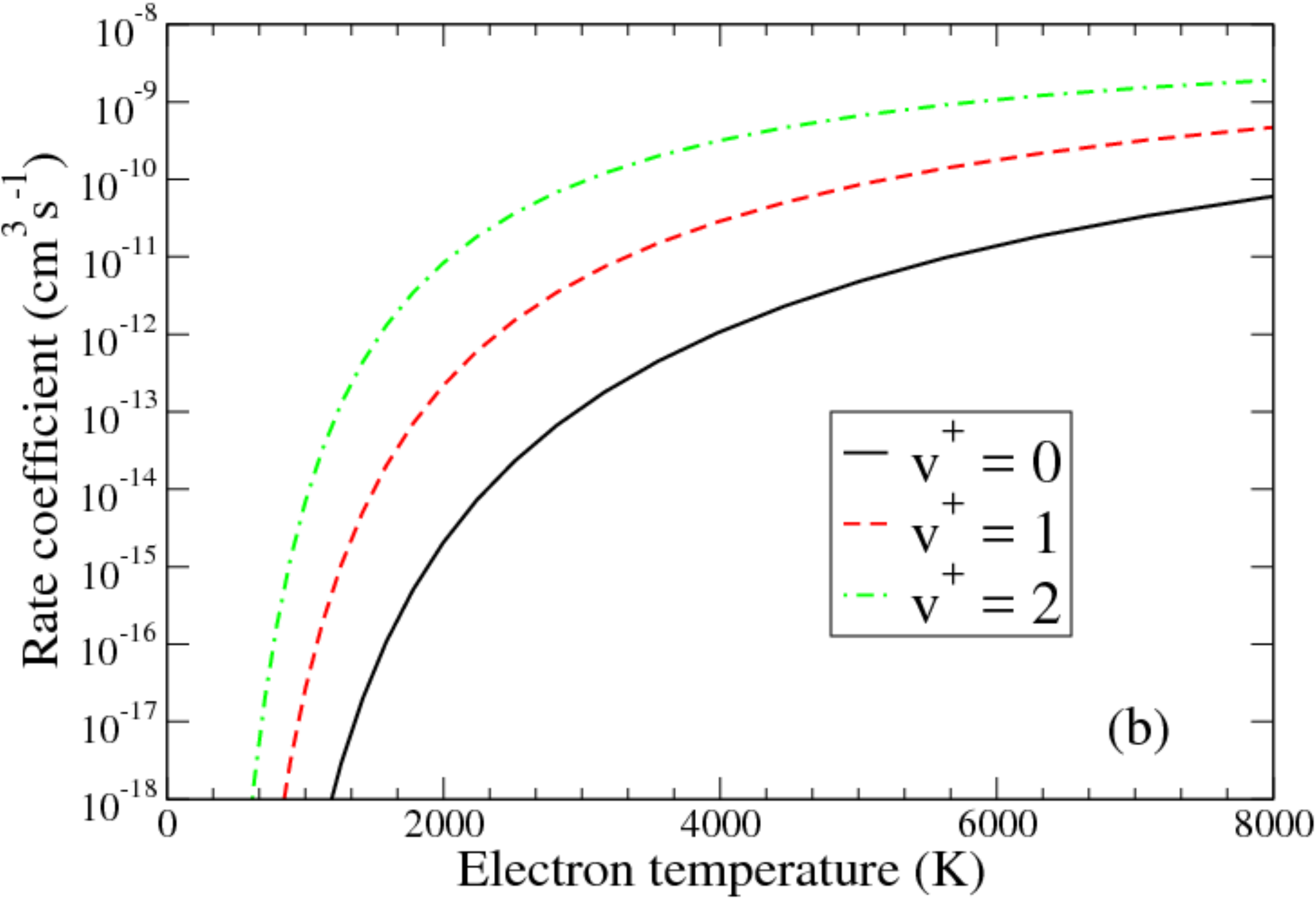}
\caption{Dissociative recombination of ArH$^+$ on its lowest vibrational levels: (a) global cross sections, coming from the sum over all the available dissociative states; (b) the corresponding Maxwellian-averaged rate coefficients. \label{fig:xsecDR}}
\end{figure*}

Figures \ref{fig:xsecDR}(a) and (b) display, respectively, the
results for DR cross sections and the corresponding rate coefficients for
$v^+=0,1,2$. Two features can be noted: 

(i) The  resonant structures present in the cross sections correspond to the temporary captures into singly-excited Rydberg states ArH$^*$, and they cease to appear when the electron energy reaches the dissociation energy of ArH$^+$($v^+=0,1,2$); 

(ii) For a vibrationally relaxed target, the dissociation channels are closed for energies of the incident electron below 1.8~eV. For the ion situated on one of the next 8 excited vibrational states, the threshold decreases progressively, and the DR becomes exothermic for vibrational levels equal or higher to 9 only. This particular energetic situation explains the particular behavior of the computed rate coefficients displayed in Fig.~\ref{fig:xsecDR}(b),  namely the very low values and the "explosive" increase below 2000 K.

\begin{figure}
\centering
\includegraphics[scale=.3]{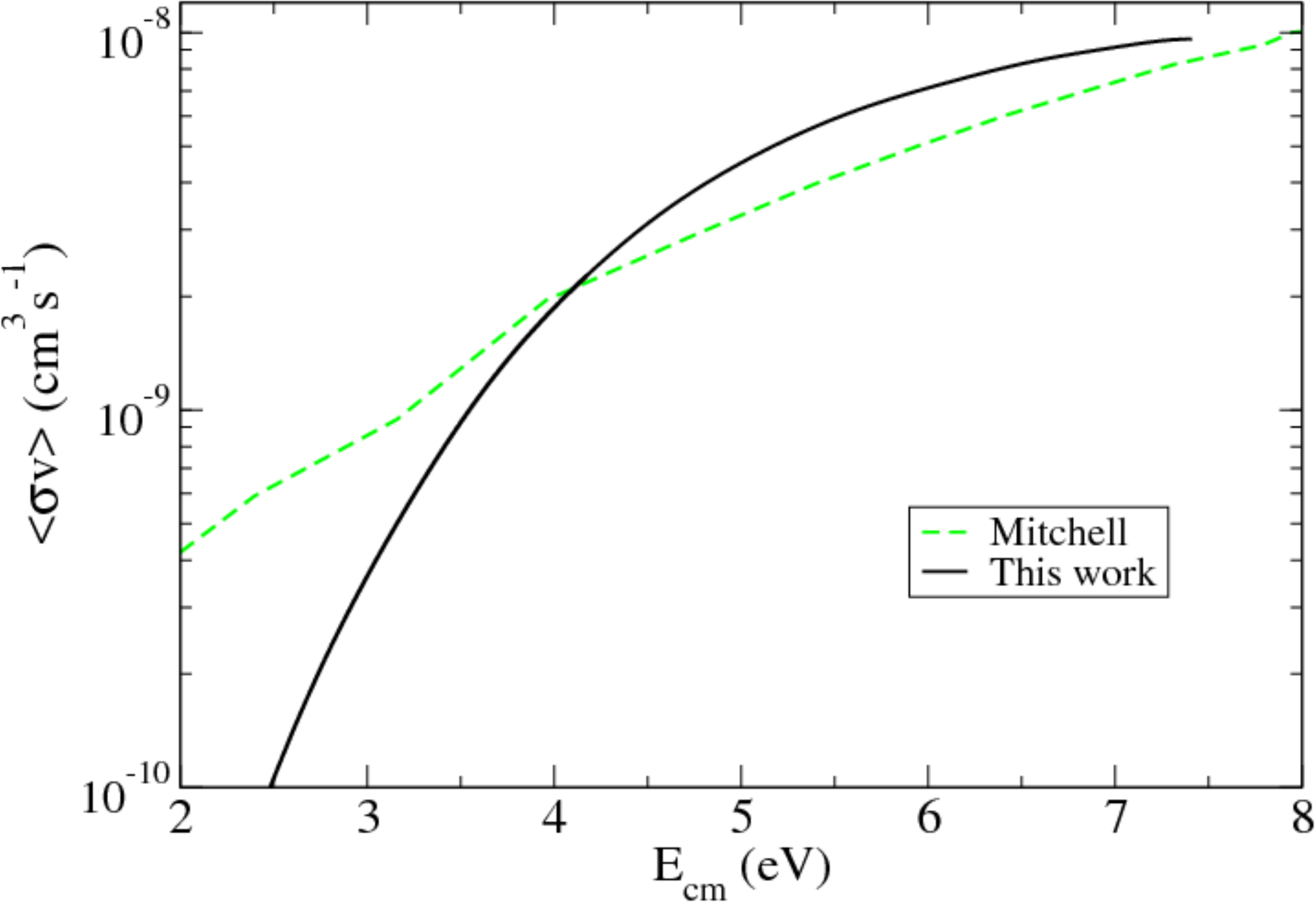}
\caption{Dissociative recombination of vibrationally relaxed  ArH$^+$. 
Comparison between the rate coefficient  measured in the CRYRING storage ring \citet{Mitchell2005}
and the anisotropic rate coefficient obtained by the convolution of our MQDT-computed cross section using the temperatures characterizing the relative velocities of the electrons with respect to the ions in the experiment. 
\label{fig:ratemitchel}}
\end{figure}

In order to validate the results, Fig.~\ref{fig:ratemitchel} shows the anisotropic DR rate coefficient for $v^+=0$, calculated by considering the electron beam with a longitudinal temperature $kT_\Vert\approx0.5$~eV and a transverse temperature $kT_\bot\approx25$~meV,
compared to the experimental data from the storage ring by \citet{Mitchell2005}. We note that the agreement is quite satisfactory at energies greater than $\sim$3 eV within the 20\% experimental error. At lower energies our calculated rates are smaller than the experimental ones: This can derive from bad detected signal as stated by the authors.

\begin{figure}
\centering
\includegraphics[scale=.3]{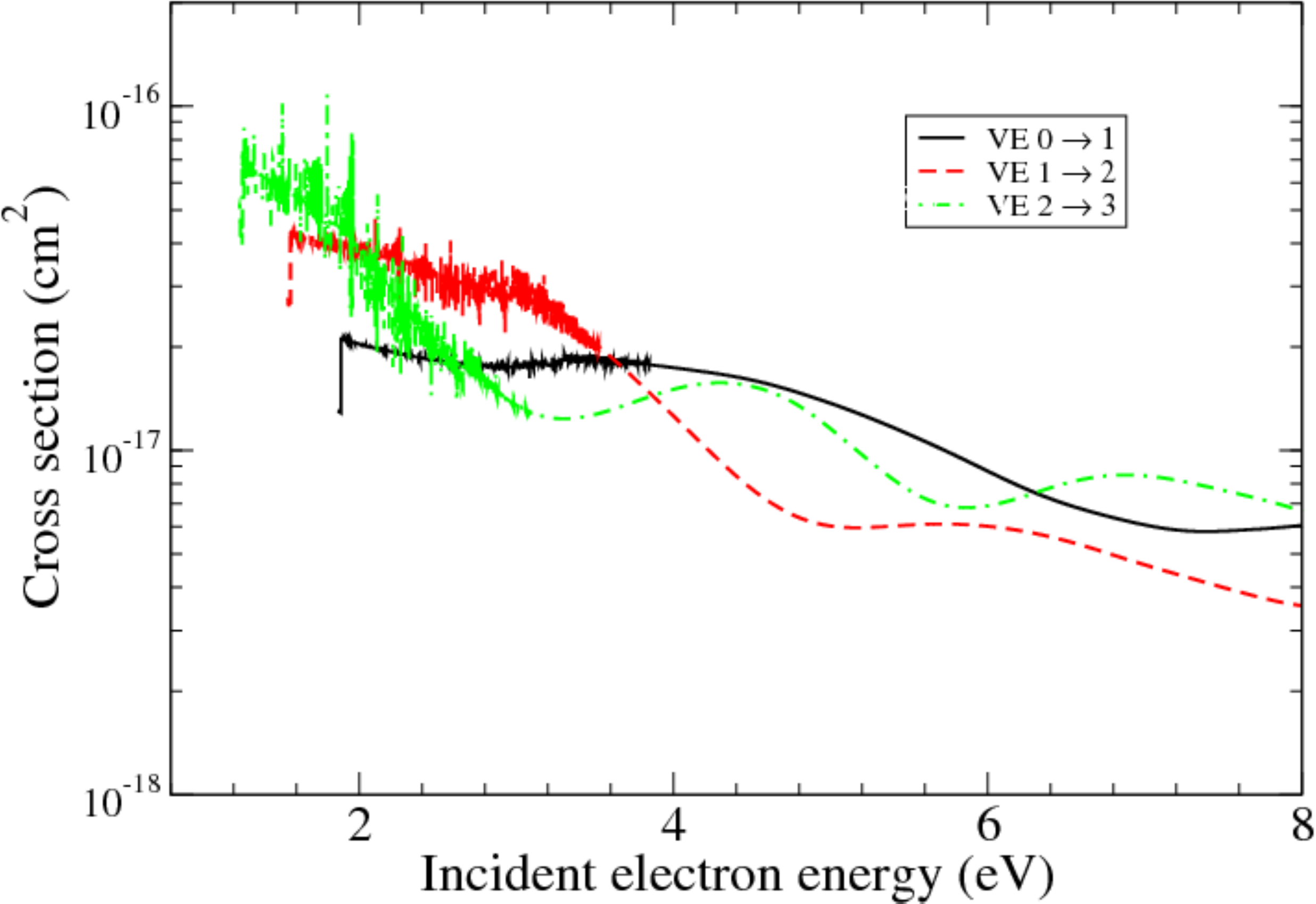}
\caption{Vibrational excitation (VE) of ArH$^+$ on its lowest vibrational levels: Cross sections for $\Delta v^+=0$ (solid lines). The dissociative recombination (DR) cross section are also shown for comparison (broken line). \label{fig:allxsec}}
\end{figure}

Figure \ref{fig:allxsec} displays the DR cross section compared to  the
competitive process of VE for one quantum excitation in the same energy range.
The main feature is that, at energies just above the opening of the dissociative channels, the VE
cross section is larger than the corresponding DR starting from the same
vibrational level. 

\begin{figure}
\centering
\includegraphics[scale=.3]{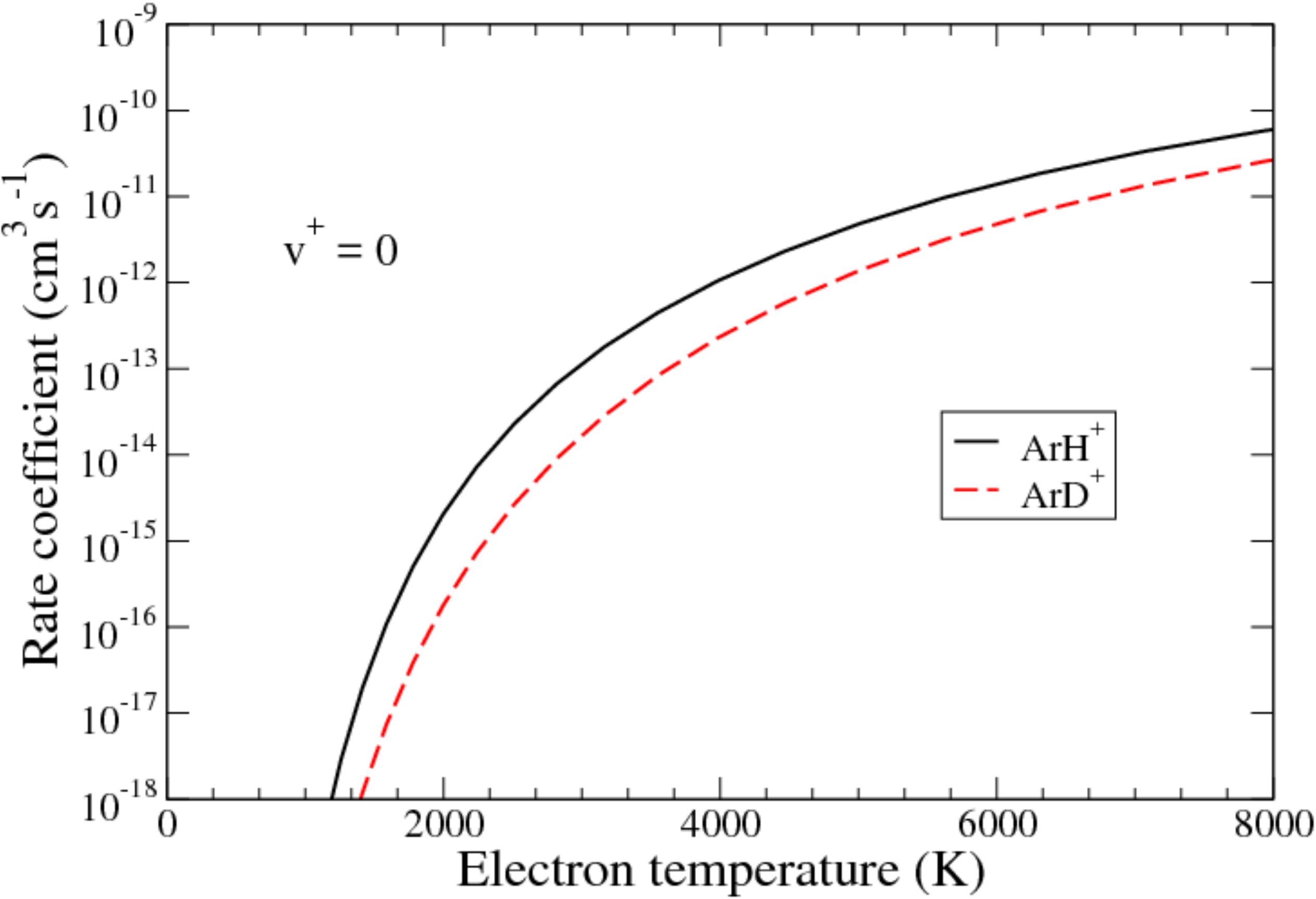}
\caption{Dissociative recombination rate of vibrationally relaxed ArH$^+$ and ArD$^+$ as a function of electron temperature: The isotopic effects. \label{fig:rateArD}}
\end{figure}

We also checked the isotopic effect by replacing ArH$^+$ by ArD$^+$, which results  in a variation of the reduced mass by a factor of 2. Fig. \ref{fig:rateArD} displays this effect for $v^+ = 0$  DR rate coefficient. The rates decrease by a factor between 10 at 1000 K and 3 at 8000 K, due to lowering of the ArD$^+$ ground state, compared to that of ArH$^+$. 

\subsection{Astrophysical consequences}

\begin{figure*}
\centering
\includegraphics[scale=.3]{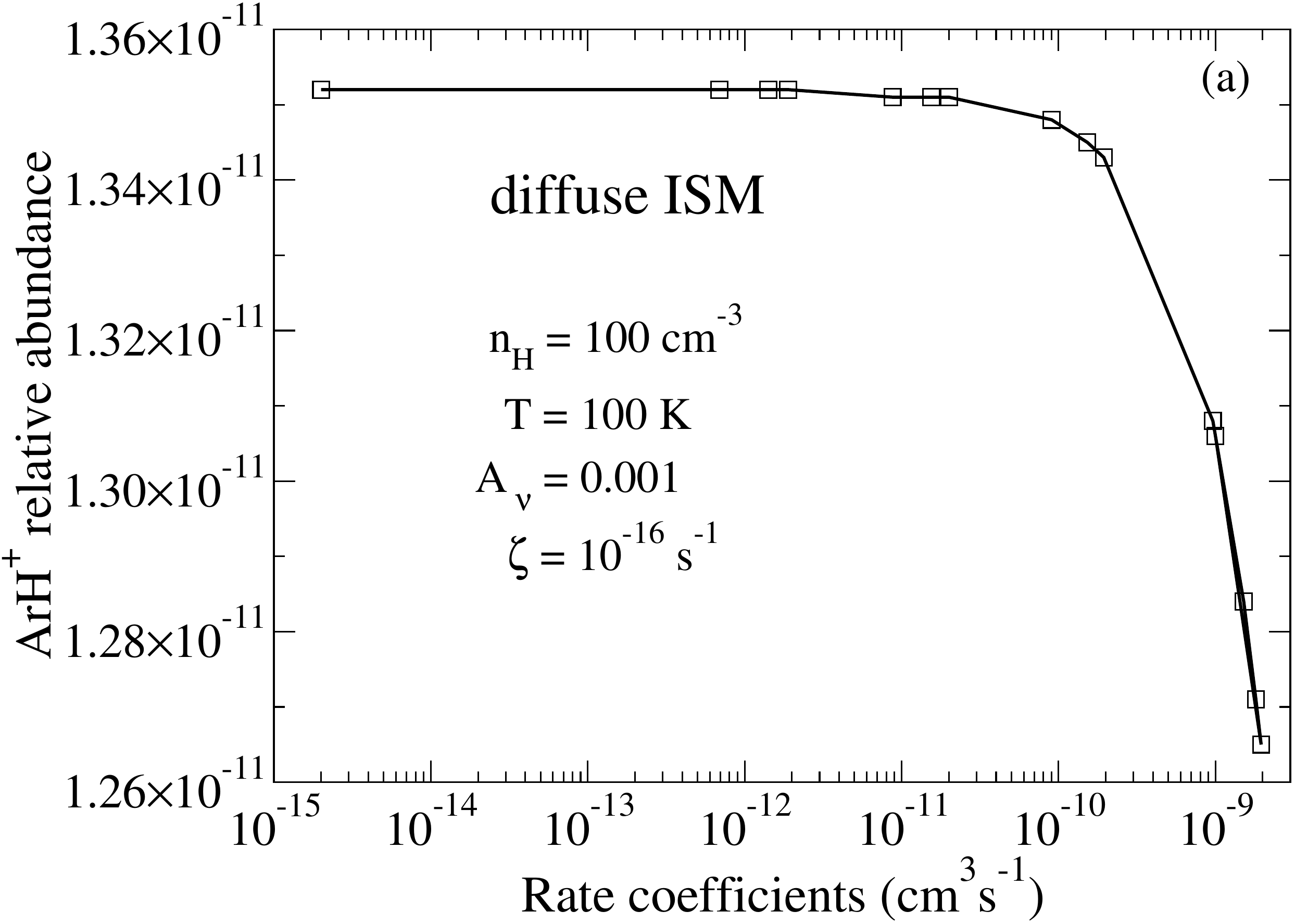} \hspace{1cm} \includegraphics[scale=.3]{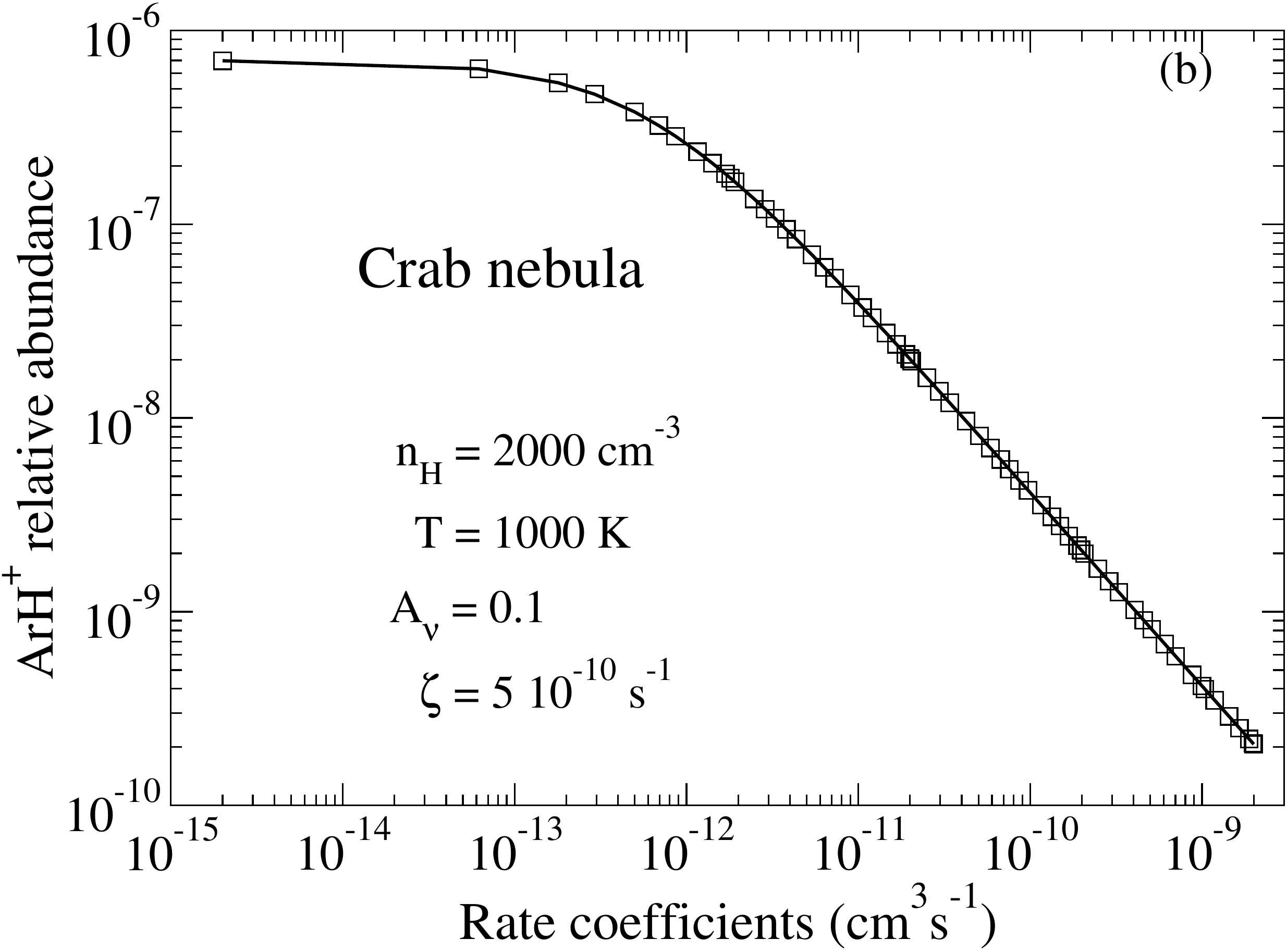}
\caption{Relative abundance of ArH$^+$ as a function of the rate coefficients for the case of (a) diffuse ISM (temperature $T=100$~K) and (b) Crab nebula (temperature $T=1000$~K). \label{fig:abundance}}
\end{figure*}

As stated previously, ArH$^+$ DR is an important destruction mechanism in interstellar conditions. We have examined two different environments where this molecular ion has been found and have varied the value of the DR rate coefficient over a range of values between $10^{-9}$ and $10^{-18}$~cm$^3$~s$^{-1}$ for a sample of 0D steady state chemical models. We solve the coupled  $\frac{d}{dt} [X] = 0 $  differential equations where [X] stands for the abundance of a particular X molecule included in the chemical network for a fixed value of density and temperature and different values of the DR chemical rate coefficient of ArH$^+$, $k_{{\rm{DR}}}$ (ArH$^+$). In Fig.~\ref{fig:abundance}(a), we display the different solutions of the argonium relative abundance as a function of $k_{\rm{DR}}$ (ArH$^+$) for typical diffuse cloud conditions: Proton density n$_{\rm{H}}=100$~cm$^{-3}$,  temperature $T =  100$~K, H$_2$ cosmic ionization rate $\zeta = 10^{-16}$~s$^{-1}$, visual extinction A$_v = 0.001$ and standard interstellar radiation field defined by the scaling parameter $\chi=1$. In Fig.~\ref{fig:abundance}(b), we display the different solutions for physical conditions pertaining to the Crab nebula, as discussed in \cite{Priestley2017}, \textit{i.e.}  n$_{\rm{H}}= 2000$~cm$^{-3}$, $T =  1000$~K, H$_2$ cosmic ionization rate $\zeta = 5$~10$^{-10}$~s$^{-1}$,  $\chi=  60$, A$_v =  0.1$ and the elemental abundances displayed in Table 1 of  \cite{Priestley2017}.  Each point corresponds to a specific model results and the line connects the different model results. In the standard diffuse cloud conditions, we see that the argonium relative abundance remains constant for values of  $k_{\rm{DR}}$ (ArH$^+$) smaller than some $10^{-11}$~cm$^3$~s$^{-1}$, where another destruction mechanism   such as photodissociation becomes dominant. It should also be noticed that the scale is linear and  the variations are moderate.
However, in the extreme conditions of the Crab nebula where the cosmic ionization rate is about 7 orders of magnitude larger, the variation of the relative fractional abundance of argonium is much more spectacular. The limiting value of $k_{\rm{DR}}$ (ArH$^+$) = $10^{-13}$~cm$^3$~s$^{-1}$,  below which the relative abundance of argonium remains almost stable and the destruction by photodissociation and reaction with H$_2$ take over the dissociative recombination.
Our theoretical computations demonstrate that the actual value is 
significantly below the experimental upper limit  $10^{-9}$ cm$^3$ s$^{-1}$ and even below the limiting values stressed out by the models (see Fig.~\ref{fig:xsecDR}(b)).
Within these findings, we conclude that DR plays a negligible role in astrophysical media and that photodissociation and reactions with molecular hydrogen become the main destruction processes.

\section{Conclusions} \label{sec:concl}

In this paper we explored the superexcited states of ArH within the R-matrix approach and we 
computed the cross sections and the corresponding rate coefficients for the dissociative recombination and the vibrational excitation of ArH$^+$ by using Multichannel Quantum Defect Theory. The very low values of the dissociative recombination rate coefficients leads to the conclusion that the only significant ArH$^+$ destruction mechanisms in the interstellar medium are the collisions with H$_2$ molecules and the photodissociation.

\section*{Acknowledgements}

ER, IFS and VL acknowledge the Programme National ``Physique et Chimie du Milieu Interstellaire'' (PCMI) of CNRS/INSU with INC/INP co-funded by CEA and CNES. They also thank for generous financial support from La R\'egion Haute-Normandie \textit{via}  the GRR Electronique, Energie et Mat\'eriaux, from the ``F\'ed\'eration de Recherche Energie, Propulsion, Environnement", and from the LabEx EMC$^3$ and FEDER \textit{via} the projects PicoLIBS (ANR-10-LABEX-09-01), EMoPlaF and CO$_2$-VIRIDIS. IFS and VL thank PHC GALILEE 2018 PROJET (39379SF) and the GdR THEMS. IFS and JZM acknowledge support from the IAEA \textit{via} the Coordinated Research Project ``Light Element Atom, Molecule and Radical Behaviour in the Divertor and Edge Plasma Regions". JZM acknowledges support from USPC \textit{via} ENUMPP and Labex SEAM. This work is supported by BATTUTA Project (Building Academic Ties Towards Universities through Training Activities) in the frame of the Erasmus Mundus program, at LOMC UMR-CNRS-6294 of Le Havre University. YM thanks the SRI department, especially Mrs. Martine Currie, for outstanding hospitality.




\bibliographystyle{mnras}

\begin{thebibliography}{}
\makeatletter
\relax
\def\mn@urlcharsother{\let\do\@makeother \do\$\do\&\do\#\do\^\do\_\do\%\do\~}
\def\mn@doi{\begingroup\mn@urlcharsother \@ifnextchar [ {\mn@doi@}
  {\mn@doi@[]}}
\def\mn@doi@[#1]#2{\def\@tempa{#1}\ifx\@tempa\@empty \href
  {http://dx.doi.org/#2} {doi:#2}\else \href {http://dx.doi.org/#2} {#1}\fi
  \endgroup}
\def\mn@eprint#1#2{\mn@eprint@#1:#2::\@nil}
\def\mn@eprint@arXiv#1{\href {http://arxiv.org/abs/#1} {{\tt arXiv:#1}}}
\def\mn@eprint@dblp#1{\href {http://dblp.uni-trier.de/rec/bibtex/#1.xml}
  {dblp:#1}}
\def\mn@eprint@#1:#2:#3:#4\@nil{\def\@tempa {#1}\def\@tempb {#2}\def\@tempc
  {#3}\ifx \@tempc \@empty \let \@tempc \@tempb \let \@tempb \@tempa \fi \ifx
  \@tempb \@empty \def\@tempb {arXiv}\fi \@ifundefined
  {mn@eprint@\@tempb}{\@tempb:\@tempc}{\expandafter \expandafter \csname
  mn@eprint@\@tempb\endcsname \expandafter{\@tempc}}}

\bibitem[\protect\citeauthoryear{Alekseyev, Liebermann  \& Buenker}{Alekseyev
  et~al.}{2007}]{Alekseyev2007}
Alekseyev A.~B.,  Liebermann H.,   Buenker R.~J.,  2007, \mn@doi [Physical
  Chemistry Chemical Physics] {10.1039/b706670h}, 9, 5088

\bibitem[\protect\citeauthoryear{Barlow et~al.,}{Barlow
  et~al.}{2013}]{Barlow2013}
Barlow M.~J.,  et~al., 2013, \mn@doi [Science] {10.1126/science.1243582}, 342,
  1343

\bibitem[\protect\citeauthoryear{Carr et~al.,}{Carr et~al.}{2012}]{Carr2012}
Carr J.~M.,  et~al., 2012, \mn@doi [The European Physical Journal D]
  {10.1140/epjd/e2011-20653-6}, 66, 58

\bibitem[\protect\citeauthoryear{Chakrabarti et~al.,}{Chakrabarti
  et~al.}{2013}]{Chakrabarti-PRA-2013}
Chakrabarti K.,  et~al., 2013, \mn@doi [Phys. Rev. A]
  {http://dx.doi.org/10.1103/PhysRevA.87.022702}, 87, 022702

\bibitem[\protect\citeauthoryear{{Ep\'ee Ep\'ee}, Mezei, Motapon, Pop  \&
  Schneider}{{Ep\'ee Ep\'ee} et~al.}{2015}]{Epee-MNRAS-2015}
{Ep\'ee Ep\'ee} M.~D.,  Mezei J.~Z.,  Motapon O.,  Pop N.,   Schneider I.~F.,
  2015, \mn@doi [MNRAS] {http://dx.doi.org/10.1093/mnras/stv2329}, 455, 276

\bibitem[\protect\citeauthoryear{Faure, Gorfinkiel, Morgan  \& Tennyson}{Faure
  et~al.}{2002}]{Faure2002224}
Faure A.,  Gorfinkiel J.~D.,  Morgan L.~A.,   Tennyson J.,  2002, \mn@doi
  [Computer Physics Communications]
  {http://dx.doi.org/10.1016/S0010-4655(02)00141-8}, 144, 224

\bibitem[\protect\citeauthoryear{Giusti}{Giusti}{1980}]{0022-3700-13-19-025}
Giusti A.,  1980, Journal of Physics B: Atomic and Molecular Physics, 13, 3867

\bibitem[\protect\citeauthoryear{Guberman \& Giusti-Suzor}{Guberman \&
  Giusti-Suzor}{1991}]{doi:10.1063/1.460913}
Guberman S.~L.,  Giusti-Suzor A.,  1991, \mn@doi [The Journal of Chemical
  Physics] {10.1063/1.460913}, 95, 2602

\bibitem[\protect\citeauthoryear{Hamilton, Faure  \& Tennyson}{Hamilton
  et~al.}{2016}]{HamiltonFaureTennyson2016}
Hamilton J.~R.,  Faure A.,   Tennyson J.,  2016, \mn@doi [Monthly Notices of
  the Royal Astronomical Society] {10.1093/mnras/stv2429}, 455, 3281

\bibitem[\protect\citeauthoryear{Hotop, Roth, Ruf  \& Yencha}{Hotop
  et~al.}{1998}]{Hotop1998}
Hotop H.,  Roth T.~E.,  Ruf M.-W.,   Yencha A.~J.,  1998, \mn@doi [Theoretical
  Chemistry Accounts] {10.1007/s002140050364}, 100, 36

\bibitem[\protect\citeauthoryear{Jungen, Roche  \& Arif}{Jungen
  et~al.}{1997}]{Jungen1481}
Jungen C.,  Roche A.~L.,   Arif M.,  1997, \mn@doi [Philosophical Transactions
  of the Royal Society of London A: Mathematical, Physical and Engineering
  Sciences] {10.1098/rsta.1997.0072}, 355, 1481

\bibitem[\protect\citeauthoryear{Kirrander, Child  \& Stolyarov}{Kirrander
  et~al.}{2006}]{B511864F}
Kirrander A.,  Child M.~S.,   Stolyarov A.~V.,  2006, \mn@doi [Phys. Chem.
  Chem. Phys.] {10.1039/B511864F}, 8, 247

\bibitem[\protect\citeauthoryear{Kramida, {Yu.~Ralchenko}, Reader  \& {and NIST
  ASD Team}}{Kramida et~al.}{2018}]{NIST_ASD}
Kramida A.,  {Yu.~Ralchenko} Reader J.,   {and NIST ASD Team} 2018, NIST Atomic
  Spectra Database (ver. 5.5.2), {NIST Atomic Spectra Database (ver. 5.5.2),
  [Online]. Available: {\tt{https://physics.nist.gov/asd}} [2018, January 30].
  National Institute of Standards and Technology, Gaithersburg, MD.}

\bibitem[\protect\citeauthoryear{Little \& Tennyson}{Little \&
  Tennyson}{2014}]{0953-4075-47-10-105204}
Little D.~A.,  Tennyson J.,  2014, Journal of Physics B: Atomic, Molecular and
  Optical Physics, 47, 105204

\bibitem[\protect\citeauthoryear{Little, Chakrabarti, Mezei, Schneider  \&
  Tennyson}{Little et~al.}{2014}]{Little-PRA-2014}
Little D.~A.,  Chakrabarti K.,  Mezei J.~Z.,  Schneider I.~F.,   Tennyson J.,
  2014, \mn@doi [PHYSICAL REVIEW A]
  {http://dx.doi.org/10.1103/PhysRevA.90.052705}, 90, 052705

\bibitem[\protect\citeauthoryear{Mitchell et~al.,}{Mitchell
  et~al.}{2005}]{Mitchell2005}
Mitchell J. B.~A.,  et~al., 2005, Journal of Physics B: Atomic, Molecular and
  Optical Physics, 38, L175

\bibitem[\protect\citeauthoryear{Motapon et~al.,}{Motapon
  et~al.}{2014}]{PhysRevA.90.012706}
Motapon O.,  et~al., 2014, \mn@doi [Phys. Rev. A] {10.1103/PhysRevA.90.012706},
  90, 012706

\bibitem[\protect\citeauthoryear{M\"{u}ller et~al.,}{M\"{u}ller
  et~al.}{2015}]{Muller2015}
M\"{u}ller H. S.~P.,  et~al., 2015, \mn@doi [Astronomy {\&} Astrophysics]
  {10.1051/0004-6361/201527254}, 582, L4

\bibitem[\protect\citeauthoryear{Neufeld \& Wolfire}{Neufeld \&
  Wolfire}{2016}]{Neufeld2016}
Neufeld D.~A.,  Wolfire M.~G.,  2016, \mn@doi [The Astrophysical Journal]
  {10.3847/0004-637x/826/2/183}, 826, 183

\bibitem[\protect\citeauthoryear{Priestley, Barlow  \& Viti}{Priestley
  et~al.}{2017}]{Priestley2017}
Priestley F.~D.,  Barlow M.~J.,   Viti S.,  2017, \mn@doi [Monthly Notices of
  the Royal Astronomical Society] {10.1093/mnras/stx2327}, 472, 4444

\bibitem[\protect\citeauthoryear{Roueff, Alekseyev  \& Bourlot}{Roueff
  et~al.}{2014}]{Roueff2014}
Roueff E.,  Alekseyev A.~B.,   Bourlot J.~L.,  2014, \mn@doi [Astronomy {\&}
  Astrophysics] {10.1051/0004-6361/201423652}, 566, A30

\bibitem[\protect\citeauthoryear{Schilke et~al.,}{Schilke
  et~al.}{2014}]{Schilke2014}
Schilke P.,  et~al., 2014, \mn@doi [Astronomy {\&} Astrophysics]
  {10.1051/0004-6361/201423727}, 566, A29

\bibitem[\protect\citeauthoryear{Stolyarov \& Child}{Stolyarov \&
  Child}{2005}]{B501400J}
Stolyarov A.~V.,  Child M.~S.,  2005, \mn@doi [Phys. Chem. Chem. Phys.]
  {10.1039/B501400J}, 7, 2259

\bibitem[\protect\citeauthoryear{Tennyson}{Tennyson}{1996}]{0953-4075-29-24-024}
Tennyson J.,  1996, Journal of Physics B: Atomic, Molecular and Optical
  Physics, 29, 6185

\bibitem[\protect\citeauthoryear{Tennyson}{Tennyson}{2010}]{Tennyson_PR_2010}
Tennyson J.,  2010, \mn@doi [Physics Reports] {DOI:
  10.1016/j.physrep.2010.02.001}, 491, 29

\bibitem[\protect\citeauthoryear{Tennyson \& Noble}{Tennyson \&
  Noble}{1984}]{Tennyson1984421}
Tennyson J.,  Noble C.~J.,  1984, \mn@doi [Computer Physics Communications]
  {10.1016/0010-4655(84)90147-4}, 33, 421

\makeatother
\end{thebibliography}



\appendix

\section{Some extra material}

The numerical data for ArH$^+$ dissociative recombination rate coefficients corresponding to the Fig.~\ref{fig:xsecDR}(b) can be found as supplementary material to this paper.


\bsp	
\label{lastpage}
\end{document}